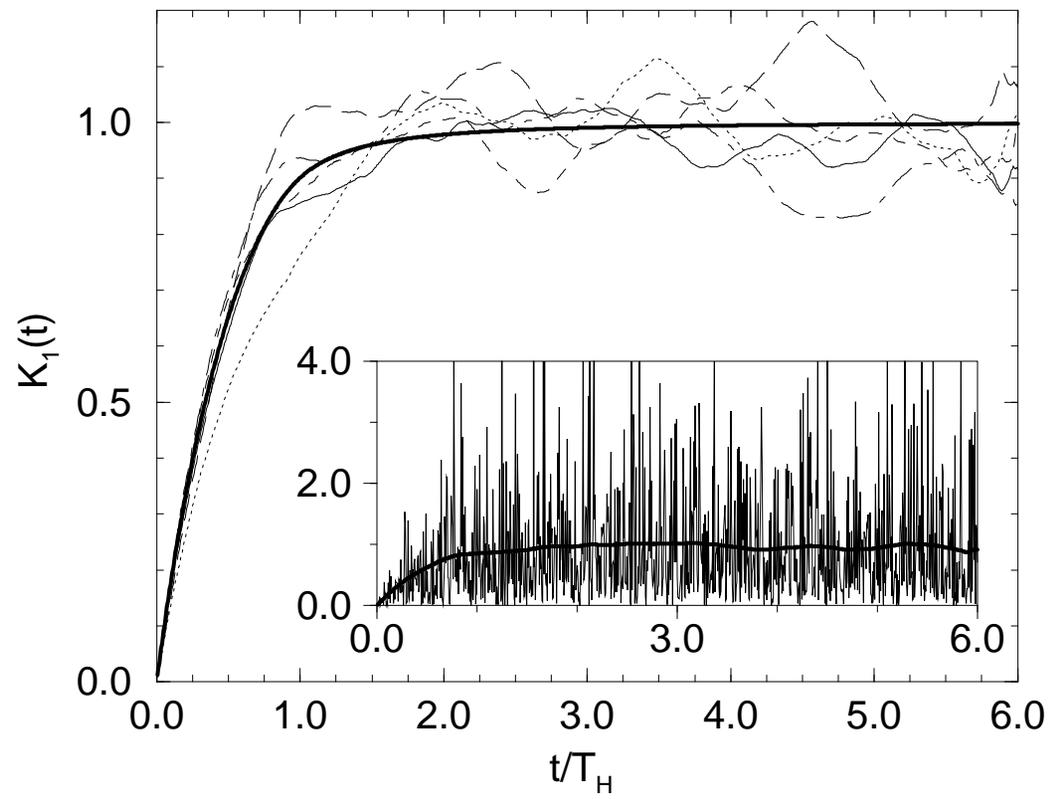

Figure 1

Figure 2

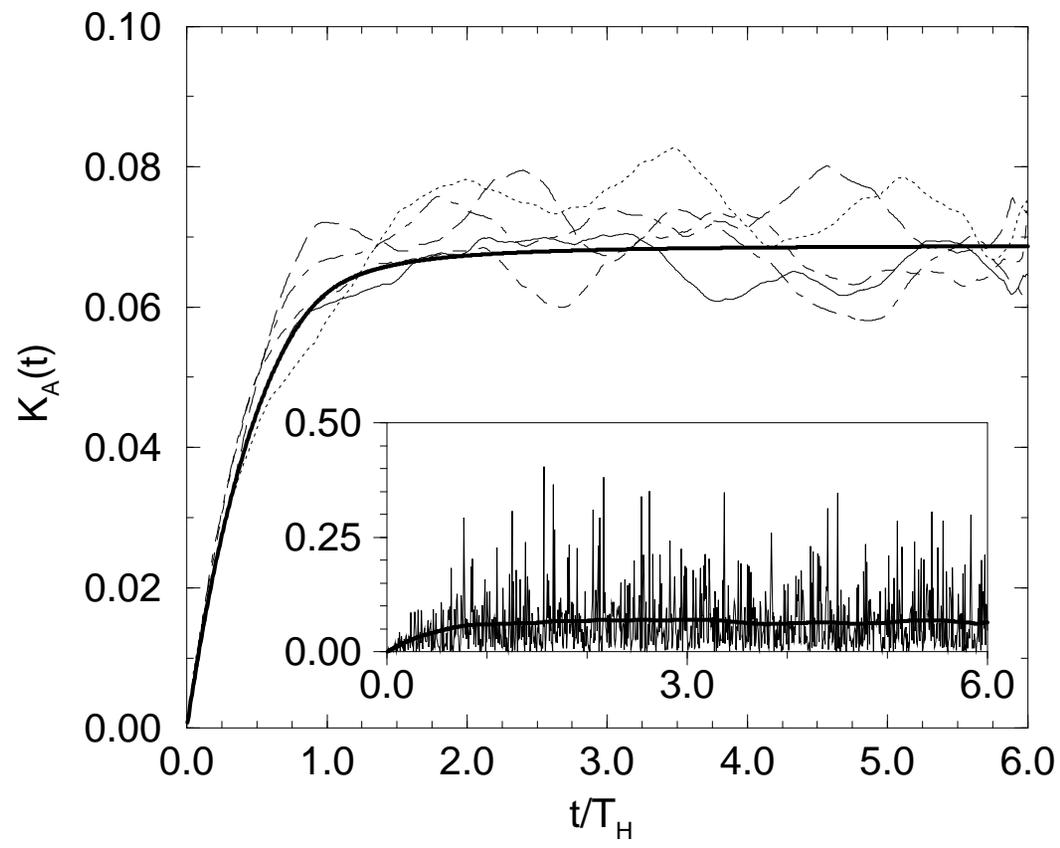

# Figure 3

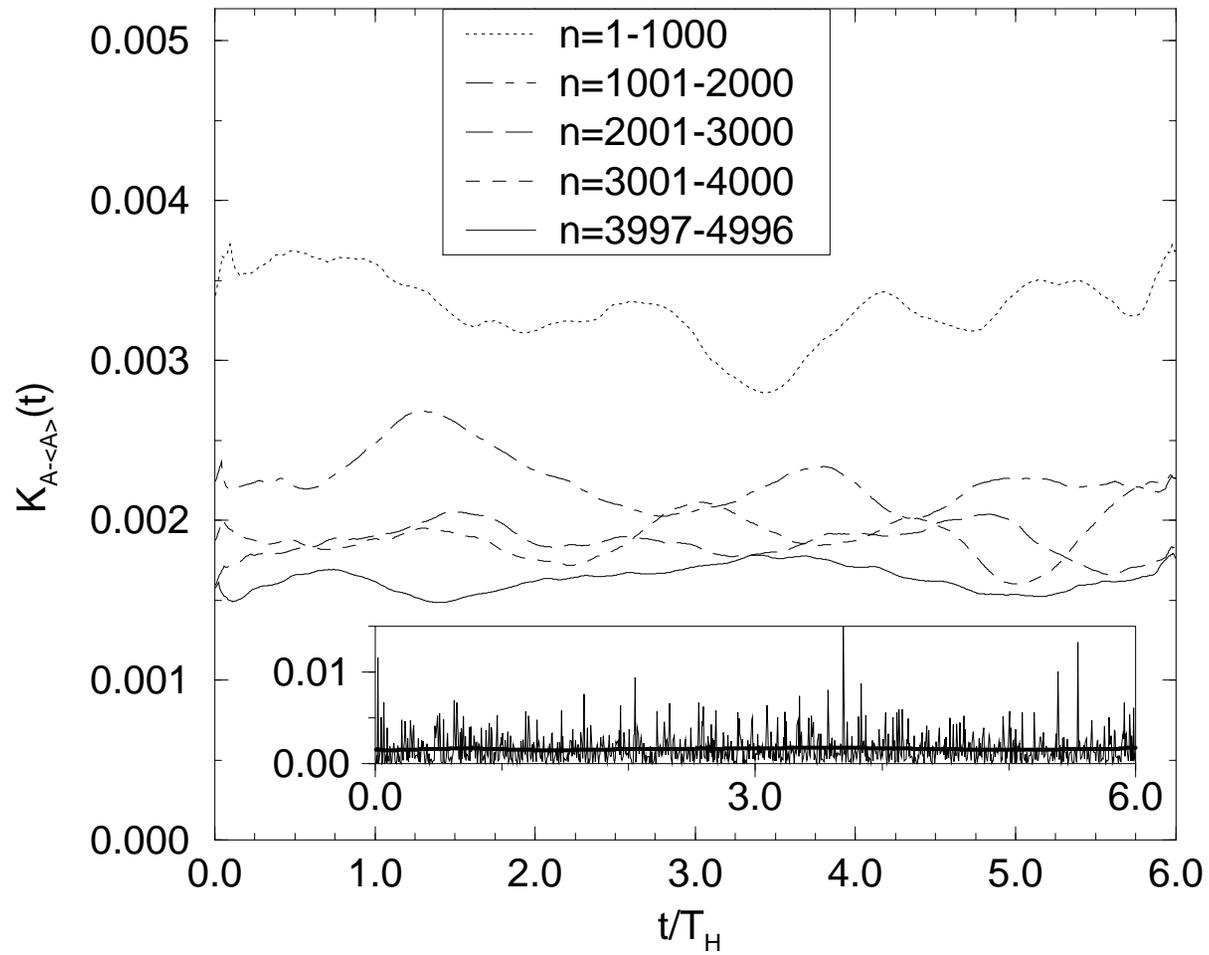

# Semiclasssical form factor of matrix element fluctuations

Bruno Eckhardt

*FB Physik und Institut für Chemie und Biologie des Meeres, Carl von Ossietzky Universität,*
*Postfach 25 03, D-26111 Oldenburg, Germany*

Jörg Main

*Institut für Theoretische Physik, Ruhr-Universität Bochum, 44780 Bochum, Germany*

(August 17, 1995)

We analyze within a semiclassical approximation the form factor for the fluctuations of quantum matrix elements around their classical average. We find two contributions: one is proportional to the form factor for the density of states, with an amplitude determined by the squared average of the matrix elements. The other is constant and related to the fluctuations of finite time classical trajectory segments around the phase space average. The results are illustrated for an observable in the quadratic Zeeman effect.

05.45.+b, 03.65.Sq

Much of the relation between classical chaos and signatures in quantum mechanics has been clarified by a combined study of semiclassical and random matrix theories [1,2]. Semiclassical analysis relates the mean of eigenvalues and matrix elements to a classical phase space volume part, the Thomas-Fermi or Weyl term and the fluctuations around the mean to periodic orbit contributions through the Gutzwiller trace formula and [3]. Random matrix theory focusses on the universal properties of the fluctuations and has predictions for Gaussian ensembles with orthogonal, unitary and symplectic symmetries, which are in good agreement with numerical observations on chaotic systems [2,4]. Berry [5] has connected both approaches in his analysis of the spectral form factor, the Fourier transform of the correlation function for the density of states. For short times, the form factor is dominated by isolated periodic orbits [6,7]. For intermediate and long times qualitative aspects of the form factor can be explained by semiclassical and quantum sum rules. A numerical demonstration of this relation will be included below.

Our main purpose here is to analyze the form factor for matrix element fluctuations. Previous semiclassical analyses have focussed on the mean behaviour [8–10] and correlations in billiard wave functions [11], periodic orbit effects in wave functions [12,13] and periodic orbit correlations in fluctuation properties [14,15]. Random matrix theory predicts Gaussian fluctuations of individual matrix elements and thus Porter-Thomas distributions for transition strengths [2,4]. One question to be answered here is that on the semiclassical prediction for the form factor of a matrix element weighted density of states.

Closely related is the problem of the size of quantum fluctuations around the classical Thomas-Fermi terms. Calculations based on this classical expression have proven rather succesful in molecular physics, but their accuracy must be taken as accidental since there have been no a priori estimates for the error. The present analysis will yield such an estimate and will quantify the expectation that the classical behaviour is recovered as $\hbar \to 0$. We will focus here on strongly chaotic systems, leaving the case of mixed phase space and further investigations to a more detailed report [16].

We consider a spinless particle in a bounded domain, with quantum eigenvalues $E_n$ ordered in an increasing sequence and eigenstates $|n\rangle$, and some observable $\hat{A}$, which will be assumed to be sufficiently smooth so that the classical limit $A(\mathbf{p}, \mathbf{q})$ exists and that the periodic orbit expressions of [15] can be applied. We will assume that the classical system is fully ergodic without any islands of stability. A fixed energy around a reference energy $E_0$ will contain an increasing number of states as $E_0 \to \infty$, while the classical mechanics changes rather little for sufficiently small intervals. We then consider the density of states, weighted by the matrix elements $\langle n|\hat{A}|n\rangle$, and projected onto an interval of width $\Delta E$ around energy $E_0$ by a window function,

$$\rho_A(E) = \sum_n \langle n|\hat{A}|n\rangle \, \delta(E - E_n) w(E_n) \,. \quad (1)$$

For the theoretical analysis we use a Gaussian window

$$w(E) = \frac{1}{\sqrt{2\pi}\Delta E} e^{-(E-E_0)^2/2\Delta E^2} \,; \quad (2)$$

the choice of window functions of course does not affect the final result but some of the prefactors in intermediate expressions. Our analysis concerns the auto-correlation function of the fluctuations of the density of states around its average $\overline{\rho}_A$ and its Fourier transform, the form factor. The auto-correlation function, defined by

$$\tilde{C}_A(\epsilon) = \int dE \, \rho_A(E + \epsilon/2)\rho_A(E - \epsilon/2) - \overline{\rho}_A^2 \,, \quad (3)$$

splits into a diagonal and an off-diagonal part,

$$\tilde{C}_A(\epsilon) = \sum_n |\langle n|\hat{A}|n\rangle|^2 w(E_n)^2 \delta(\epsilon)$$
$$+ \sum_{n \neq m} \langle n|\hat{A}|n\rangle \langle m|\hat{A}|m\rangle w(E_n) w(E_m) \delta(\epsilon - (E_n - E_m)) \,. \quad (4)$$

A convenient normalization is to divide by the weighted number of states under the window,

$$C_N = \sum_n w(E_n)^2 \approx \frac{\overline{\rho}\Delta E}{2\sqrt{\pi}}, \qquad (5)$$

where $\overline{\rho}$ is the mean density of states; the remaining factors are specific for the Gaussian window. With this normalization, the strength of the delta function at the origin in the normalized function

$$C_A(\epsilon) = \tilde{C}_A(\epsilon)/C_N \qquad (6)$$

becomes 1 for the density of states and the average of the square of the matrix elements $\langle n|\hat{A}|n\rangle$ in the general case.

The form factor is the Fourier transform of the autocorrelation function,

$$K(t) = \int d\epsilon\, C(\epsilon) \exp(i\epsilon t/\hbar) \qquad (7)$$

Since the $\delta$-function in $C(\epsilon)$ is mapped into the large $t$ behaviour of $K(t)$, the density of states and the size of matrix elements can be read off from the large $t$ behaviour of the form factor.

The semiclassical analysis starts from the Gutzwiller trace formula [3,17,15],

$$\rho_{A,sc}(E) = \overline{\rho}_A + \frac{1}{2\pi\hbar}\sum_p w_p A_p e^{iT_p(E-E_0)/\hbar}, \qquad (8)$$

where the first term is the smooth phase space average of the observable $A$ and where the sum extends over all periodic orbits and multiple traversals. This form assumes that the window $\Delta E$ around $E_0$ is sufficiently small so that the actions can be expanded linearly in energy, giving rise to the exponential in (8). The term

$$A_p = \int dt\, A(\mathbf{p}(t), \mathbf{q}(t)) \qquad (9)$$

denotes the integral of the observable along the periodic orbit $p$. The remaining term $w_p$ depends on the monodromy matrix, phases due to the action at energy $E_0$ and the Maslov indices. The semiclassical approximation for the form factor of the matrix element weighted density of states then is the absolute value squared of the Fourier transform of the product of (8) with the window function, divided by the normalization factor $C_N$. Upon division by the normalization $C_N$, and after cancellation of various prefactors related to the window function one finds within the diagonal approximation [5]

$$K_{A,sc} \approx \frac{1}{T_H}\sum_p |w_p|^2 A_p^2 \delta_\tau(t-T_p), \qquad (10)$$

where $T_H = 2\pi\overline{\rho}\hbar$ is the Heisenberg time.

The widths of the $\delta$-function is related to the widths of the initial window, $\tau = \hbar/\Delta E$. For short times $t$, the $\delta$-functions do not overlap, and one can identify individual periodic orbits [6,7]. For longer times, the $\delta$-functions overlap, forming a quasi-continuum. Since the amplitude of the orbits contributing to this quasi-continuum is given by purely classical quantities, one can evaluated its amplitude in hyperbolic systems under the assumption that the periodic orbits are densely distributed in phase space [18–21]. In particular, for the density of states with $\hat{A} = 1$, the integrals $A_p$ become the periods $T_p$ and one can use the classical sum rule [18–21]

$$\sum_p |w_p|^2 T_p^2 \delta_\tau(t-T_p) = \langle T_p^2\rangle_t \sim t \qquad (11)$$

to obtain

$$K_{1,sc} \sim g\, t/T_H \qquad (12)$$

where the factor $g = 2$ for orthogonal systems and $g = 1$ for unitary systems is related to the presence or absence of unitary symmetries [5].

If the observable is not the identity, the result depends on the fluctuation properties of the classical integrals $A_p$. Since long trajectories will ergodically sample phase space, one can expect that on average $A_p \sim \overline{A} T_p$, where $\overline{A}$ denotes the classical ergodic average of the observable $A$. Assuming that correlations along trajectories decay sufficiently rapidly, as in hyperbolic systems, the distribution around this average will be Gaussian,

$$P_T(A)dA = \text{Prob}\ \{A_p \in [A, A+dA] \text{ and } T_p \text{ near } T\}$$
$$= \frac{1}{\sqrt{2\pi\alpha T}}\exp-(A-\overline{A}T)^2/\alpha T, \qquad (13)$$

where the variance increases linearly with the period $T$ of the orbits. Thus the average of the squares of $A_p$ near a period $T$ becomes

$$\langle A_p^2\rangle \sim \overline{A}^2 T^2 + \alpha T. \qquad (14)$$

Combining this with the classical periodic orbit sum rule (11), one arrives at the diagonal approximation for the form factor

$$K_{A,sc}(t) \sim g\overline{A}^2 t/T_H + g\alpha/T_H \qquad (t \ll T_H). \qquad (15)$$

Thus the form factor for matrix elements contains in addition to the linearly increasing part a constant off-set. This off-set is determined by the fluctuation properties of averages of the observable computed from classical trajectory segment.

For longer times, one can appeal to the behaviour of the quantum correlation function [5] and argue that $K(t)$ has to settle at a constant, related to the strength of the delta function in $C(\epsilon)$. In a simple approximation one can assume that the semiclassical form is correct up to a time $t_c \sim T_H/g$ and that $K(t)$ keeps the value thus attained also for larger $t$. For the density of states form factor, this assumption yields the correct expression for the Gaussian unitary ensemble. Since uniformly hyperbolic systems seem to fall into the universality class of Gaussian random matrix models [2], we will below compare our results to the functional form of the random matrix theory form factor.

The asymptotic amplitude relates matrix elements and classical phase space averages,

$$\overline{\langle n|\hat{A}|n\rangle^2} \sim g\overline{A}^2 + g\alpha/T_H . \quad (16)$$

The first term is to be expected from the classical phase space average. The second term is a correction present for finite densities of states (finite $T_H$) only, and can be calculated from the fluctuations of classical trajectory segments. If the average of the matrix elements is subtracted, only this term survives and it then measures the variance of the quantum matrix elements. It varies like $1/T_H$ and vanishes as $\hbar \to 0$. In non-hyperbolic systems, the dependence on $T_H$ may be different, as explained in [16].

We test these predictions for eigenstates for hydrogen in a magnetic field (for reviews, see [22,23]). The periodic orbit contributions to the density of states have been studied in [15]. Due to scaling, the classical dynamics depends on a combination of energy $E$ and scaled magnetic field $\gamma = B/B_0$ with $B_0 = 2.35 \cdot 10^5\, T$ only. At scaled energy $\epsilon = E\gamma^{-2/3} = -0.1$, a large fraction of phase space is chaotic and the small elliptic islands that will still be present are not felt on the time scale relevant for the present analysis. In this sense, the system is hyperbolic. A numerical solution of the Schrödinger equation for quantizing magnetic field strengths $z_n = \gamma_n^{-1/3}$ in semi-parabolic coordinates produced 4996 converged eigenstates and matrix elements for the observable $A = 1/2r$, the potential energy. For the numerical determination of the form factor, these eigenstates were divided into five lots of 1000 states each, modulated by a Hamming window [24]. The Fourier transforms were calculated in the density of states as a functions of the scaling variable $z$, the corresponding Heisenberg time being $T_H = 2\pi/\Delta z$, where $\Delta z$ is the mean spacing of the eigenvalues.

Fig. 1 shows the form factor of the density of states and compares it to the form factor for the Gaussian orthogonal ensemble from random matrix theory,

$$K_{GOE}(t) = \begin{cases} 2t - t\ln(1+2t) & (t \leq 1) \\ 2 - t\ln((2t+1)/(2t-1)) & (t \geq 1) \end{cases} . \quad (17)$$

To eliminate the large fluctuations evident in the inset, the quotient of the calculated Fourier transform and the GOE form factor was averaged over an interval $[\max\{0, t-0.6\}, \min\{t_{max}, t+0.6\}]$, weighted with a Parzen window [24]. The fitted curve thus is

$$K_{fit}(t) = \left\langle \frac{K_{calc}(t)}{K_{GOE}(t)} \right\rangle K_{GOE}(t) . \quad (18)$$

The remaining fluctuations indicate the size of the deviations to be expected.

The form factor for the matrix element weighted density of states is shown in Fig. 2. Since the average of the matrix elements is non-zero, one can still see the linear increase. The constant term to be discussed below is smaller than the fluctuations, so that the overall form is again compatible with that of a GOE form factor. The asymptotic amplitude (0.068) compares favourably with the square of the average matrix elements ($0.26^2 = 0.0676$).

To highlight the second contribution to (16) from the fluctuations of the classical trajectory segments, we show in Fig. 3 the form factor for the fluctuations of the marix elements minus their average. Within the fluctuations, this function is constant, as expected. As the quantum numbers and $T_H$ increase, the constant decreases. For the Heisenberg times relevant to Fig. 3 there are still some correlations and the rate of decay is not $1/T_H$, but the size of the fluctuations is still given by the variance of classical trajectory segments [16].

In summary, a semiclassical analysis of the form factor of the matrix element weighted density of states has revealed a dominant term due to the phase space average of the classical observable and a second contribution due to the fluctuations of classical trajectory segments. The shape of the first term is found to be qualitatively similar to form factor for the density of states (semiclassical theory does at present not allow one to go further). The second part opens the possibility of estimate from classical trajectories not just the average of matrix elements but also the size of the quantum fluctuations around this average, with important consequences for the classical trajectory calculations in atomic and molecular physics [25]. Further discussions and examples, as well as an extension to non-hyperbolic systems, will be discussed elsewhere [16].

*Acknowledgments* It is a pleasure to thank J. Keating, S. Fishman for discussions and K. Müller for classical calculations.

[1] B. Eckhardt, Phys. Rep. **163**, 205 (1988)
[2] O. Bohigas, Chaos and Quantum Physics, Les Houches Lecture Notes, Session LII, M.J. Giannoni, A. Voros and J. Zinn-Justin (eds), North Holland, Amsterdam, 1991, p. 87
[3] M. C. Gutzwiller, Chaos in Classical and Quantum Mechanics, Springer (New York 1990).
[4] M.L. Mehta, Random Matrices, Academic, London, 1991
[5] M.V. Berry, Proc. R. Soc. (London) **A 400**, 229 (1985)
[6] R.Balian and C. Bloch, Ann. Phys. **85**, 514 (1974)
[7] D. Wintgen, Phys. Rev. Lett. **58**, 1589 (1987)
[8] A. Voros, Ann. Inst. Poincar'e, **26**, 343 (1977)

[9] M. Feingold and A. Peres, Phys. Rev. A **34**, 591 (1986)
[10] T. Prosen and M. Robnik, J. Phys. A **26**, L319 (1993)
[11] M.V. Berry, J. Phys. A **10**, 2083 (1977)
[12] E.J. Heller, Phys. Rev. Lett. **53**, 1515 (1984)
[13] E.B. Bogomolny, Physica D **31**, 169 (1988)
[14] M. Wilkinson, J. Phys. A **20**, 2415 (1987), ibid **21**, 1173 (1988)
[15] B. Eckhardt, S. Fishman, K. Müller and D. Wintgen, Phys. Rev. A **45** 3531 (1992)
[16] B. Eckhardt, S. Fishman, J. Keating, O. Agam, J. Main and K. Müller, in preparation
[17] B. Eckhardt, Proceedings of the International School of Physics "Enrico Fermi", Course CXIX "Quantum Chaos", G. Casati, I. Guarneri, and U. Smilansky (eds), North Holland, Amsterdam, 1993, p. 77.
[18] L.P. Kadanoff and C. Tang, Proc. Natl. Acad. Sci USA **81**, 1276 (1984)
[19] P. Cvitanović and B. Eckhardt, J. Phys. A **24**, L237 (1991)
[20] J.H. Hannay and A.M. Ozorio de Almeida, J. Phys. A **17**, 3429 (1984)
[21] N. Argaman, Y. Imry and U. Smilansky, Phys. Rev. B **47**, 4440 (1993)
[22] H. Friedrich and D. Wintgen, Phys. Rep. **183**, 37 (1989)
[23] D. Delande, Chaos and Quantum Physics, Les Houches Lecture Notes, Session LII, M.J. Giannoni, A. Voros and J. Zinn-Justin (eds), North Holland, Amsterdam, 1991, p. 665
[24] W.H. Press et al., Numerical Recipies, Cambridge 1990
[25] J.M. Gomez-Llorente and E. Pollak, Ann. Rev. Phys. Chem. **43**, 91 (1992)

FIG. 1. Spectral form factor $K_1(t)$ for the eigenvalues of hydrogen in a magnetic field at scaled energy $\epsilon = -0.1$. The different curves correspond to the eigenvalues $1 - 1000$ (dotted), $1001 - 2000$ (dash-dotted), $2001 - 3000$ (long dashed), $3001 - 4000$ (short dashes) and $3997 - 4996$ (continuous). For comparison, a fit to the GOE form factor as determined from (17) is also shown (thick line). The inset shows the original, fluctuating data for the eigenvalues $n = 3997 \ldots 4996$.

FIG. 2. Form factor $K_A(t)$ for the eigenvalues of hydrogen in a magnetic field weighted with the matrix elements of $\hat{A} = 1/2r$. The GOE form factor shown for comparision was adjusted by the asymptotic amplitude. Otherwise same as Fig. 1.

FIG. 3. Form factor $K_A(t)$ for the eigenvalues of hydrogen in a magnetic field weighted with the matrix elements of $\hat{A} = 1/2r - \overline{A}$. Since now the average of the matrix elements vanishes, there is no linear increase, the form factor is constant. The original data are shown in the inset, the averages were computed with a Parzen window.